\documentclass[aps,twocolumn,longbibliography]{revtex4-1}

\usepackage{amsmath,amssymb}
\usepackage{graphicx}
\usepackage[]{hyperref}

\newcommand{\rb}{\textsuperscript{87}Rb}
\newcommand{\ent}{$\mathcal{E}_\mathrm{Ent}$}
\newcommand{\eprab}{$\mathcal{E}_\mathrm{EPR}^{A \rightarrow B}$}
\newcommand{\eprba}{$\mathcal{E}_\mathrm{EPR}^{B \rightarrow A}$}
\newcommand{\heia}{$\mathcal{E}_\mathrm{Hei}^A$}
\newcommand{\heib}{$\mathcal{E}_\mathrm{Hei}^B$}

\newcommand{\ket}[1]{\left\vert #1 \right\rangle}

\begin{document}
\title{Einstein-Podolsky-Rosen experiment with two Bose-Einstein condensates} 

\author{Paolo Colciaghi}
\altaffiliation{These authors contributed equally to this work}
\affiliation{Department of Physics, University of Basel, 4056 Basel, Switzerland}
\author{Yifan Li}
\altaffiliation{These authors contributed equally to this work}
\affiliation{Department of Physics, University of Basel, 4056 Basel, Switzerland}
\author{Philipp Treutlein}
\affiliation{Department of Physics, University of Basel, 4056 Basel, Switzerland}
\author{Tilman Zibold}
\email{tilman.zibold@unibas.ch}
\affiliation{Department of Physics, University of Basel, 4056 Basel, Switzerland}

\begin{abstract}
In 1935, Einstein, Podolsky and Rosen (EPR) conceived a Gedankenexperiment which became a cornerstone of quantum technology and still challenges our understanding of reality and locality today. While the experiment has been realized with small quantum systems, a demonstration of the EPR paradox with spatially separated, massive many-particle systems has so far remained elusive. We observe the EPR paradox in an experiment with two spatially separated Bose-Einstein condensates containing about 700 Rubidium atoms each. EPR entanglement in conjunction with individual manipulation of the two condensates on the quantum level, as demonstrated here, constitutes an important resource for quantum metrology and information processing with many-particle systems. Our results show that the conflict between quantum mechanics and local realism does not disappear as the system size is increased to over a thousand massive particles.
\end{abstract}

\maketitle

\section{Introduction}
According to quantum mechanics, complementary properties of a physical system such as its position and momentum or two orthogonal components of its spin cannot be simultaneously known with arbitrary precision. This is expressed by the Heisenberg uncertainty principle \cite{heisenberg_1927,heisenberg_1930}, which imposes a lower bound on the product of the two uncertainties, so that better knowledge of one property necessitates greater uncertainty of the other and vice versa. This is in stark contrast to classical physics, where all properties of a system can be simultaneously known, in principle with arbitrary precision. 

In their seminal work \cite{einstein_1935}, EPR considered this aspect for a bipartite quantum system where the parts $A$ and $B$ are entangled through interactions. After spatially splitting the system, measurements on the two parts yield strongly correlated outcomes, which allow one to use measurements on $A$ to predict properties of $B$ (and vice versa). The uncertainty product of such predictions for two complementary properties of B falls below the Heisenberg uncertainty bound of $B$ \cite{reid_1989}. The fact that the distant system A can be used for better predictions than what is locally possible at B revealed a conflict between quantum mechanics and the classical principle of local realism, which became known as the ``EPR paradox'' \cite{schroedinger_1935}. In a local realist world, each system possesses its properties independent of observation and independent of actions performed on spatially separated systems. In a quantum world, on the other hand, the measurements on $A$ change the quantum state of the distant system $B$, a scenario Schr\"odinger called ``steering'' \cite{schroedinger_1935}. It was later found that not all entangled states are able to show such strong correlations \cite{werner_1989,wiseman_2007}. Only a strict subset, the so-called EPR entangled states, are able to demonstrate an EPR paradox \cite{reid_2009}. Furthermore, EPR entanglement was identified as a resource for quantum technologies such as quantum metrology, quantum teleportation, entanglement swapping, or randomness certification \cite{cavalcanti_2017b,uola_2020,yadin_2021}. 

The EPR paradox has been observed with small systems of few photons or atoms, in its original form \cite{ou_1992,hagley_1997,bowen_2003,reid_2009,kheruntsyan_2012} or in the form of Bell tests\cite{freedman_1972,lamehirachti_1976,aspect_1982,hensen_2015,giustina_2015,shalm_2015,rosenfeld_2017}. How far quantum behavior extends into the macroscopic world is an open question \cite{leggett_2002}, which can be addressed by performing EPR experiments with increasingly macroscopic, massive systems \cite{leggett_2007,cavalcanti_2007,reid_2009}. Generating and verifying sufficiently strong entanglement between massive many-particle systems is a challenging task, requiring excellent isolation from the environment, high fidelity coherent manipulations and low noise detection. Previous experiments demonstrated entanglement (non-separability) between spatially separated atomic ensembles \cite{julsgaard_2001,chou_2005}, mechanical oscillators \cite{kotler_2021,lepinay_2021}, and hybrid systems \cite{lettner_2011,thomas_2020}. However, the observed correlations were not strong enough to demonstrate an EPR paradox. 

Atomic Bose-Einstein condensates (BECs) are many-particle systems that are particularly well suited to investigate non-classical phenomena at the quantum-to-classical boundary \cite{pezze_2018}. Being composed of neutral atoms in ultra-high vacuum, they couple very weakly to their environment and thus show excellent coherence. Furthermore, they are reliably initialized in pure states and can be manipulated and detected with high fidelity by means of radio-frequency, microwave and optical fields. Multi-particle entanglement in BECs has been demonstrated in the form of squeezed spin states \cite{pezze_2018}. These states are EPR entangled, which has been verified by measuring spin correlations within a single cloud of atoms \cite{peise_2015,fadel_2018,kunkel_2018,lange_2018}, making them a promising starting point for an EPR experiment.

\begin{figure}
    \includegraphics[scale=1]{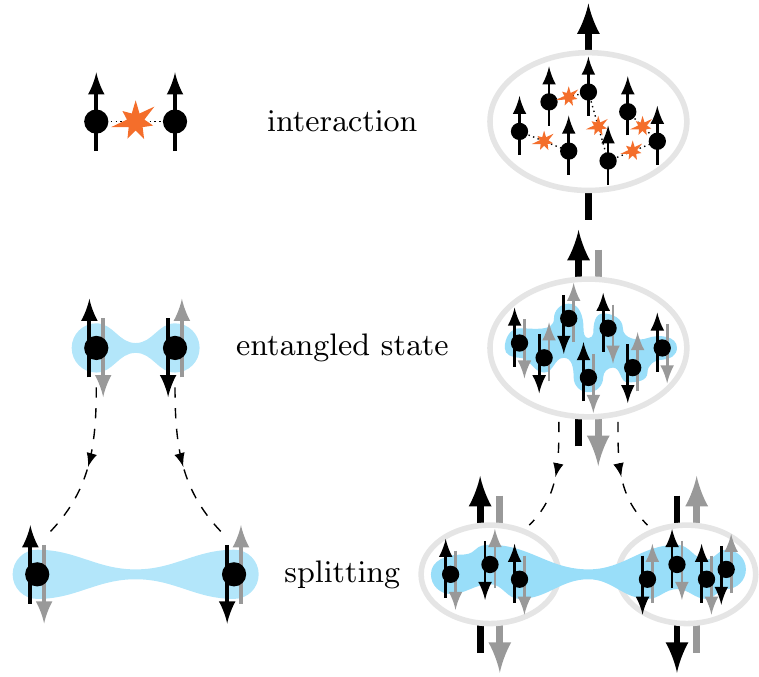}
    \caption{
        Schematic of an EPR experiment with two particles (left) and with two many-particle systems (right), where the spin degree of freedom is considered. In both cases, the particles are entangled by interactions and subsequently split into two different locations. In the case of the many-particle system, the interactions produce multipartite entanglement, which is inherited by the split systems in form of bipartite entanglement between their collective spins.
    }
    \label{many_body_EPR}
\end{figure}

In this work we perform an EPR experiment with two BECs, in close analogy with the original Gedankenexperiment, see Fig.~\ref{many_body_EPR}: We first prepare a BEC in a squeezed spin state, in which all atoms are entangled with each other. We then physically split it into two distinct condensates, which can be individually manipulated and detected on the quantum level. The two condensates inherit the entanglement from the initial state, resulting in correlated measurement outcomes and allowing us to demonstrate the EPR paradox between two massive many-particle systems.

\section{Experimental sequence}
\begin{figure*}
    \includegraphics[width=\textwidth]{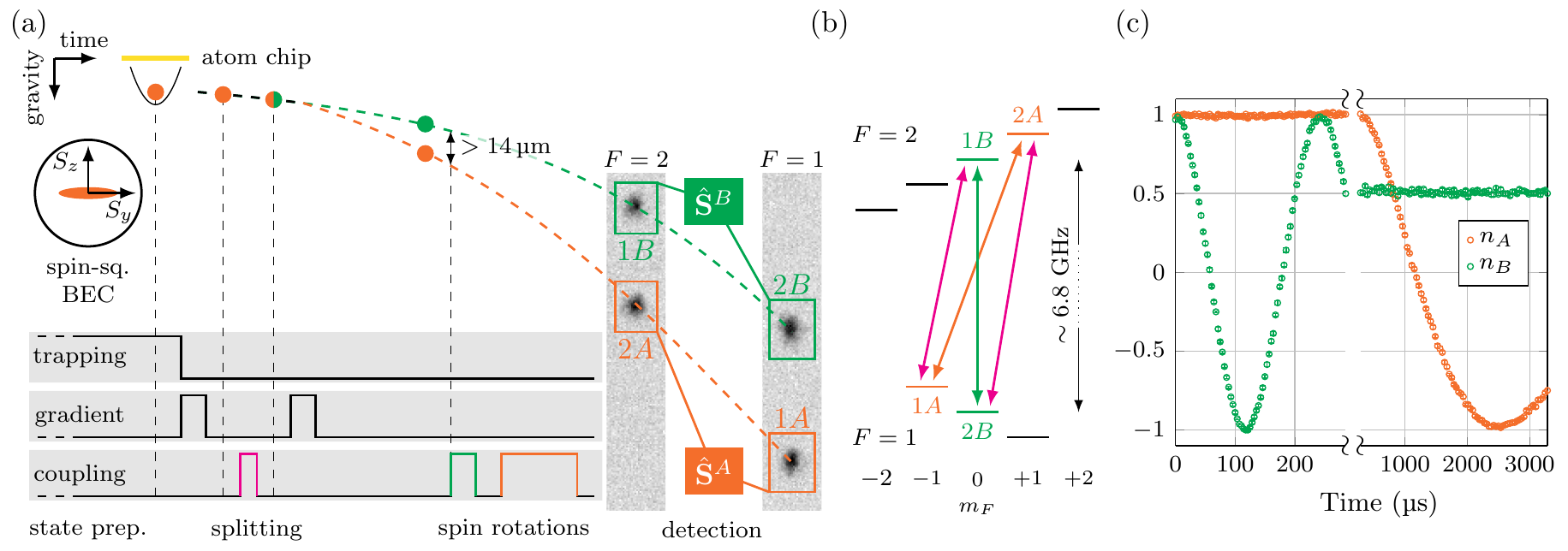}
    \caption{
        Experimental sequence for spatial splitting and independent coherent control of the two condensates.
        (a) Space-time schematic.
        Orange and green color refer to condensate A and B, respectively, whereas magenta represents the splitting pulse.
        Bottom left: main control parameters for splitting and state manipulation. The timing is aligned with the sketch above, but not to scale.
        On the right, typical absorption images are shown, from which the atom numbers in all four states are determined. This corresponds to a measurement of the selected components of the collective spins $\hat{\mathbf{S}}^A$ and $\hat{\mathbf{S}}^B$.
        (b) Hyperfine levels of the $^{87}$Rb ground state with transitions for coherent splitting and manipulation of $\hat{\mathbf{S}}^A$ and $\hat{\mathbf{S}}^B$.
        (c) Individual Rabi oscillations of the two collective spins after spatial splitting. Shown are the normalized spin components ${n_{A,B} = (N_1^{A,B} - N_2^{A,B}) / (N_1^{A,B} + N_2^{A,B})}$.
        The discontinuity in the horizontal axis denotes a change in time scale.
    }
    \label{experiment}
\end{figure*}

Our experiment starts by preparing a two-component \rb{} BEC of approximately 1400 atoms in a static magnetic trap on an atom chip \cite{bohi_2009}.
The two components refer to the internal degree of freedom: The atoms occupy the hyperfine levels ${\ket{1A} \equiv \ket{F=1,m_F=-1}}$ and ${\ket{2A} \equiv \ket{F=2,m_F=1}}$ of the electronic ground state, see Fig.~\ref{experiment}(b), forming a collection of pseudo-spin $1/2$ particles.
Due to the nearly identical collisional interaction strength and magnetic moments of these states, the atoms occupy a single spatial mode.
Hence, we can describe the BEC as a collective spin $\hat{\mathbf{S}}$, the sum of the individual atomic spins \cite{pezze_2018}.
For instance, the $ z $ component of the collective spin is given by half the population difference between the two states, $ {\hat{S}_z = (\hat{N}_1 - \hat{N}_2) / 2} $, which we can directly measure.
By coherently driving the two-photon transition between $\ket{1A}$ and $\ket{2A}$ with radio frequency and microwave fields (orange in Fig.~\ref{experiment}(b)), we are able to perform arbitrary spin rotations with high fidelity.

We create entanglement between the atomic spins in the initial BEC through interactions, by inducing one-axis twisting dynamics \cite{kitagawa_1993} with a state-dependent potential on the atom chip \cite{riedel_2010,ockeloen_2013,schmied_2016,fadel_2018}. After an interaction time of $\approx 40$~ms and a subsequent spin rotation, the BEC is prepared in a squeezed spin state polarized along $\hat{S}_x$ with squeezed spin component $\hat{S}_z$ (see Fig.~\ref{experiment}(a)), whose variance is reduced by ${- 7~\mathrm{dB}}$ compared to a coherent spin state \cite{pezze_2018}.

We then coherently split the spin-squeezed BEC into two spatially separated, individually addressable condensates (see Fig.~\ref{experiment}(a)).
While coherent splitting has been demonstrated in early experiments with BECs \cite{andrews_1997}, the particular challenge we face here is to spatially split a two-component condensate, while maintaining nearly perfect overlap and coherence between the components. This is necessary so that high-fidelity coherent spin rotations can be carried out after the splitting on each condensate separately.

To split the BEC, we first release it from the trap and accelerate it with a magnetic field gradient to reduce the overall expansion time. We then coherently transfer a fraction of the atoms to states with zero magnetic moment by simultaneously driving the transitions ${\ket{1A} \rightarrow \ket{1B} \equiv \ket{F=2,m_F=0}}$ and ${\ket{2A} \rightarrow \ket{2B} \equiv \ket{F=1,m_F=0}}$ with a two-tone microwave pulse (magenta in Fig.~\ref{experiment}). We choose to transfer half of the populations (pulse duration ${t_{\pi /2} \approx 70~\mathrm{\mu s}}$ for both transitions), which allows us to realize a nearly ideal 50:50 beam splitter for the atoms. A subsequent pulse of the magnetic field gradient selectively accelerates the atoms in states $\ket{1A}$ and $\ket{2A}$, spatially separating the system into two distinct two-component BECs, which we call system $A$ (composed of states $\ket{1A}$ and $\ket{2A}$) and system $B$ (composed of states $\ket{1B}$ and $\ket{2B}$). Since the overlap of the states in each system is preserved by the splitting mechanism, we can describe them as two collective spins $\hat{\mathbf{S}}^A$ and $\hat{\mathbf{S}}^B$. Both transition frequencies are insensitive to magnetic field fluctuations, ensuring long coherence times \cite{li_2020}.

Once the two condensates are split, we can coherently drive the transitions ${\ket{1A} \leftrightarrow \ket{2A}}$ and ${\ket{1B} \leftrightarrow \ket{2B}}$ with distinct radio-frequency and microwave signals, which allows us to perform arbitrary spin rotations on $\hat{\mathbf{S}}^A$ and $\hat{\mathbf{S}}^B$ independently -- as demonstrated by the individual Rabi oscillations shown in Fig.~\ref{experiment}(c). Subsequently, projective measurements of both collective spins are carried out by resonant absorption imaging: The atoms in states with $ F=2 $ ($\ket{2A}$ and $\ket{1B}$) are detected on a first image by a resonant laser pulse. On a second image we detect atoms with $F=1$ ($\ket{1A}$ and $\ket{2B}$), after they have been optically pumped to $F=2$. Due to the large separation between the two BECs ($ \approx 80~\mathrm{\mu m} $ at the time of the first image and $ \approx 100~\mathrm{\mu m} $ at the time of the second), we can count the atoms present in all four states separately and thus obtain a measurement of the spin components ${\hat{S}_z^A = (\hat{N}_1^A - \hat{N}_2^A)/2}$ and ${\hat{S}_z^B = (\hat{N}_1^B - \hat{N}_2^B)/2}$ (see Fig.~\ref{experiment}(a)). Other spin components can be measured by coherently rotating the collective spins before detection.

\section{EPR criteria}
According to quantum mechanics, $\hat{\mathbf{S}}^B$ satisfies the Heisenberg uncertainty relation ${\mathcal{E}_\mathrm{Hei}^{B} \equiv 4 ~ \mathrm{Var}(\hat{S}_z^{B}) \mathrm{Var}(\hat{S}_y^{B}) / \vert\langle \hat{S}_x^{B} \rangle\vert^2 \geq 1}$, which places a lower bound on the uncertainty product of $\hat{S}_z^{B}$ and $\hat{S}_y^{B}$. It follows from the spin commutation relations and applies to repeated measurements on identically prepared systems. 
A similar relation holds for $\hat{\mathbf{S}}^A$.
However, if the two spins are entangled, their measurement outcomes are correlated, allowing us to use measurement outcomes obtained on $\hat{\mathbf{S}}^A$ to predict those obtained on $\hat{\mathbf{S}}^B$. The accuracy of this prediction is quantified by  the inferred variances $\mathrm{Var}_\mathrm{inf}\left(\hat{S}_{z,y}^B\right) \equiv \mathrm{Var}\left(\hat{S}_{z,y}^B-\hat{S}_{z,y}^{B,\mathrm{inf}}\right)$, which involve linear estimates $\hat{S}_{z,y}^{B,\mathrm{inf}} \equiv - g_{z,y} \hat{S}_{z,y}^A + c_{z,y}$ of $\hat{S}_{z,y}^B$ using $\hat{S}_{z,y}^A$. Here, $g_{z,y}$ and $c_{z,y}$ are real numbers that can be chosen to optimize the prediction, i.e.\ to ensure that $\langle \hat{S}_{z,y}^{B,\mathrm{inf}}\rangle = \langle \hat{S}_{z,y}^B \rangle$ and to minimize the inferred variances. 

The EPR paradox is demonstrated if the product of the inferred variances of two non-commuting spin components is lower than the associated Heisenberg uncertainty bound \cite{reid_2009}, that is if the inequality 
\begin{equation}\label{E EPR}
\mathcal{E}_\mathrm{EPR}^{A \rightarrow B} \equiv \frac{4 ~ \mathrm{Var}_\mathrm{inf}\left(\hat{S}_z^B\right) \mathrm{Var}_\mathrm{inf}\left(\hat{S}_y^B\right)}{\left\vert\left\langle \hat{S}_x^B \right\rangle\right\vert^2} \geq 1
\end{equation}
is violated. 
Observing a violation of Eq.~(\ref{E EPR}) challenges our classical notions of locality and reality: it implies that an experimenter with access to a spatially separated system $\hat{\mathbf{S}}^A$ can make better predictions about $\hat{\mathbf{S}}^B$ than what is possible if one has full local control over $\hat{\mathbf{S}}^B$ alone.

A related but weaker criterion exists for entanglement (non-separability) \cite{giovannetti_2003}:
Two spins $\hat{\mathbf{S}}^A$ and $\hat{\mathbf{S}}^B$ are in an entangled state if the inequality
\begin{equation}\label{E Ent}
\mathcal{E}\mathrm{_{Ent}} \equiv \frac{4 ~ \mathrm{Var}\left(g_{z} \hat{S}_{z}^A + \hat{S}_z^B\right) \mathrm{Var}\left(g_{y} \hat{S}_{y}^A + \hat{S}_y^B\right)}{\left( \left\vert g_z g_y \right\vert \left\vert\left\langle \hat{S}_x^A \right\rangle\right\vert + \left\vert\left\langle \hat{S}_x^B \right\rangle\right\vert \right)^2} \geq 1
\end{equation}
is violated.
\eprab{} is always larger or equal than \ent{}, reflecting the fact that an observation of the EPR paradox requires stronger correlations than a demonstration of entanglement \cite{wiseman_2007}.

\section{Observation of the EPR paradox}
\begin{figure*}
    \includegraphics[scale=1]{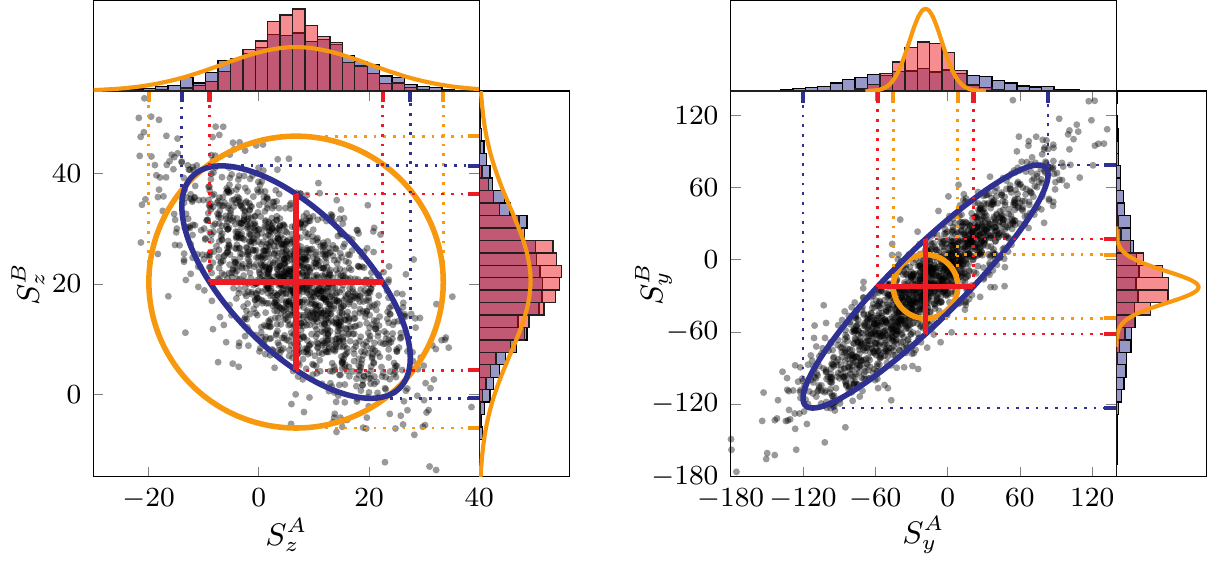}
    \caption{
        Spin correlations between the two BECs and illustration of the inferring mechanism.
        The grey dots are individual data points of simultaneous measurements of spin components $S_z^A$ and $S_z^B$ (left plot) and $S_y^A$ and $S_y^B$ (right plot) of the two systems. 
        Data points for $ S_y^B $ are corrected for the phase shift due to the measured trigger jitter of the microwave generator (Appendix~\ref{app_rf_mw_phase}).
        The blue histograms are their marginal distributions, $2\sigma$ intervals are indicated by blue dotted lines.
        The correlations of measurement results ($2\sigma$ covariance ellipses in blue) allow one to infer measurement results of one system from the other.
        This reduces the variance of the prediction as shown by the histograms and the reduced $2\sigma$ intervals in red.
        For comparison, the $2\sigma$ variance ellipses of ideal non-entangled states with the same number of atoms are shown in yellow. 
    }
    \label{spin correlations}
\end{figure*}

To evaluate these criteria, we perform measurements of either the $x$, $y$, or $z$ spin components simultaneously on both systems, repeating the experiment many times with identical preparation. The measurement basis is selected by rotating the two collective spins individually when they are completely separated, at a distance of more than $14~\mathrm{\mu m} $, see Fig.~\ref{experiment}(a). The outcome of every measurement of $ \hat{S}_y^{A,B} $ or $ \hat{S}_z^{A,B} $ (approx.\ 1600 repetitions each) is represented by a point in the correlation plots of Fig.~\ref{spin correlations}. Strong correlations are visible between $A$ and $B$ in both spin components. The inference corresponds to an affine transformation (e.g.\ $S_z^B \mapsto S_z^B + g_z {S}_z^A - c_z$), which reduces the variance, as can be seen from the marginal histograms and $2\sigma$ intervals of the raw data (blue) and the transformed quantities (red). Splitting a squeezed spin state with a beam-splitter-like process retains part of the reduced (increased) fluctuations in the squeezed (anti-squeezed) component. This can be seen from the comparison of the variance ellipse of a separable coherent spin state with the same atom number (yellow lines in Fig.~\ref{spin correlations}) to the covariance ellipse of the data (blue).

The measurements of $ \langle\hat{S}_x^{A,B}\rangle $ quantify the lengths of the collective spins $ \hat{\mathbf{S}}^{A,B} $ and thus determine the Heisenberg uncertainty bounds. Their values normalized to the atom numbers correspond to the interferometric contrast, which is a measure of the overall coherence of the process. We obtain $ {96\%} $ contrast for both $ \hat{\mathbf{S}}^A $ and $ \hat{\mathbf{S}}^B $, comparable to the contrast we measure without splitting the condensate, indicating that the reduction is mostly due to spin squeezing \cite{wineland_1994} and imperfections in the initial state preparation, not due to the splitting process.

Combining all measurements we obtain 
\begin{align*}
\mathcal{E}_\mathrm{EPR}^{A \rightarrow B} &= 0.81 \pm 0.03~,\\
\mathcal{E}_\mathrm{EPR}^{B \rightarrow A} &= 0.77\pm 0.03~,\\
\mathcal{E}_\mathrm{Ent} &= 0.35 \pm 0.02~,\\
\mathcal{E}_\mathrm{Hei}^{B} &= 10.2 \pm 0.4~,\\
\mathcal{E}_\mathrm{Hei}^{A} &= 9.2 \pm 0.5~,
\end{align*}
demonstrating both entanglement and the EPR paradox between the condensates $A$ and $B$. The Heisenberg uncertainty products \heia{} and \heib{} are larger than unity due to technical noise. The EPR criteria are much smaller due to the variance reduction by inferring. We note that all criteria are determined without subtraction of any technical noise and that we observe the paradox both ways, inferring from $A\rightarrow B$ and from $B \rightarrow A$.

\section{Individual control of the BECs}
\begin{figure}
    \includegraphics[scale=1]{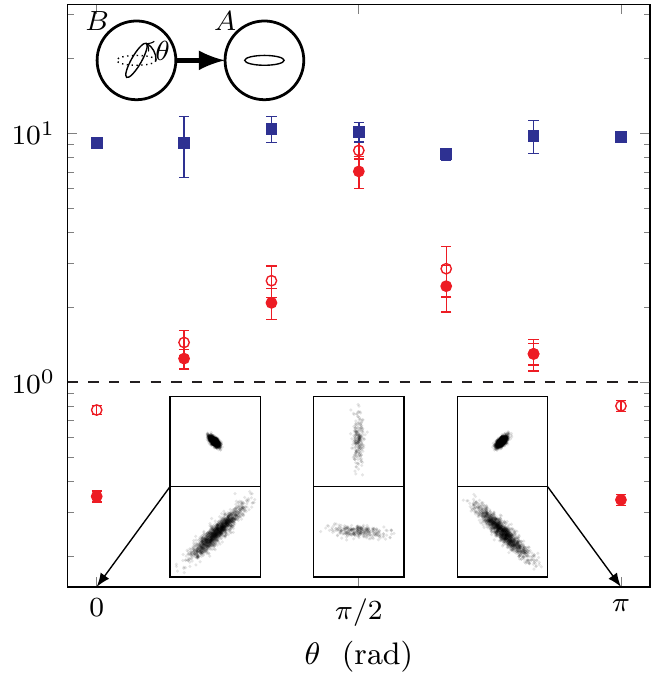}
    \caption{
        Individual manipulation of the two entangled BECs on the quantum level.
        $ \hat{\mathbf{S}}_B $ is rotated by an angle $\theta$ around the $ x $ axis with respect to $ \hat{\mathbf{S}}_A $, as sketched in the inset on the top left.
        The red filled circles represent \ent{}, red empty circles \eprba, and blue squares \heia{}.
        The EPR paradox (entanglement) is observed if \eprba{} (\ent{}) falls below the dashed line at unity. 
        The insets at the bottom show the spin correlations similar to Fig.~\ref{spin correlations}, with $\hat{S}_z$ on top and $\hat{S}_y$ at the bottom, for $\theta=0,\;\pi / 2$ and $\pi$.}
    \label{rotate B}
\end{figure}

Although we rotate $ \hat{\mathbf{S}}_A $ and $ \hat{\mathbf{S}}_B $ individually, the results presented so far were obtained by measuring the two spins in the same bases.
Since many applications of EPR entanglement require performing different measurements on the two systems, we demonstrate that our experiment is able to maintain the entanglement between $ \hat{\mathbf{S}}_A $ and $ \hat{\mathbf{S}}_B $ in this process. 
Figure~\ref{rotate B} shows the results of measurements in which $ \hat{\mathbf{S}}_B $ is rotated by an angle $\theta$ around the $x$ axis with respect to $ \hat{\mathbf{S}}_A $, as sketched in the inset on the top left.
The case $ \theta = 0 $ (leftmost points in Fig.~\ref{rotate B}) corresponds to the same configuration as above, i.e.\ simultaneous measurements of the same spin component.
As $\theta$ increases, the correlations decrease and almost completely vanish for $\theta = \pi / 2$, where orthogonal spin components are measured on the two condensates. This can be directly seen in the correlation plots at the bottom of Fig.~\ref{rotate B}. 
For $\theta= \pi$ the correlations reappear, but with opposite sign, since the spin components are now anti-aligned. The parameters \ent{} and \eprba{} fall again below unity, demonstrating that the manipulation preserves the quantum correlations.

In the setting $ \theta = \pi / 2 $ our experiment realizes a situation discussed by Schr\"odinger \cite{schroedinger_1935b}, where the values of two complementary properties of system $A$ are apparently obtained in a single experimental run: one ($\hat{S}_{z}^A$) by direct measurement on $A$ and the other ($\hat{S}_{y}^A$) by exploiting the strong correlations to infer its value from the simultaneous measurement on $B$. Under the local realist assumptions that measurements reveal pre-existing properties of a system and that simultaneous measurements on spatially separated systems do not disturb each other, the restrictions imposed by the Heisenberg uncertainty relation could thus be overcome \cite{schroedinger_1935b}. Today, however, we know that local realism is inconsistent with the results of increasingly rigorous experimental tests of Bell inequalities \cite{freedman_1972,lamehirachti_1976,aspect_1982,hensen_2015,giustina_2015,shalm_2015,rosenfeld_2017}. In the spirit of Peres' statement that ``unperformed experiments have no results'' \cite{Peres_1978}, we should thus refrain from inferring a value for $\hat{S}_{y}^A$ if it is not actually measured on system $A$.

\section{Conclusions}
Our experiment demonstrates the EPR paradox, a cornerstone of quantum physics, between two systems with a large number of massive particles. This shows that the conflict between quantum mechanics and the classical principles of locality and realism does not disappear in systems of increasing size and complexity, at least up to the level demonstrated here. The key to this result is the high degree of coherence of our splitting technique together with the ability to perform high-fidelity coherent rotations of the individual systems after splitting. 
The ability to measure the two collective spins in different bases is also an essential prerequisite for a future Bell test, which in addition requires non-Gaussian measurements or state preparation, e.g.\ single-atom resolving detection \cite{oudot_2019}. 

EPR entanglement is a valuable resource for both quantum metrology and quantum information processing.
The noise reduction gained from the inference in Eq.~\eqref{E EPR}, quantified by the difference between the Heisenberg products and the EPR criteria, translates to a metrological enhancement that can be exploited in quantum sensing \cite{yadin_2021,fadel_2022}.
One can, e.g., use one system as a small sensor to probe fields and forces with high spatial resolution and the other one as a reference to reduce the quantum noise of the first system.
Furthermore, EPR entanglement is the resource that guarantees the efficacy of certain quantum information protocols such as quantum teleportation, entanglement swapping, one-side device-independent quantum key distribution, or randomness certification \cite{reid_2009,cavalcanti_2017b,uola_2020}.
For this resource to be useful, however, the systems need to be spatially separated and individually addressable, which we realize here for the first time in massive many-particle systems.
Therefore, in addition to its foundational significance, our work demonstrates the necessary ingredients to exploit EPR entanglement in many-particle systems as a resource.

\begin{acknowledgements}
The authors acknowledge contributions to improving the experimental apparatus by Simon Josephy and Clara Piekarsky. This work was supported by the Swiss National Science Foundation (Grant No. 197230).
\end{acknowledgements}

\appendix

\section{Addressability and strength of the transitions}\label{app_address_trans}
In our experiment, multiple transitions in the hyperfine-split ground state of \rb{} are coupled with radio-frequency and microwave driving fields in order to coherently split and manipulate the two-component BEC. To preserve the entanglement in this process, it is essential to prevent loss from the system by spurious driving of undesired transitions. This requires a careful choice of parameters since several of the transitions involved are nearly degenerate and differ only by quadratic Zeeman shifts, which are of the same order of magnitude as the Rabi frequencies (1-10~kHz).

\begin{figure*}
    \includegraphics[scale=1]{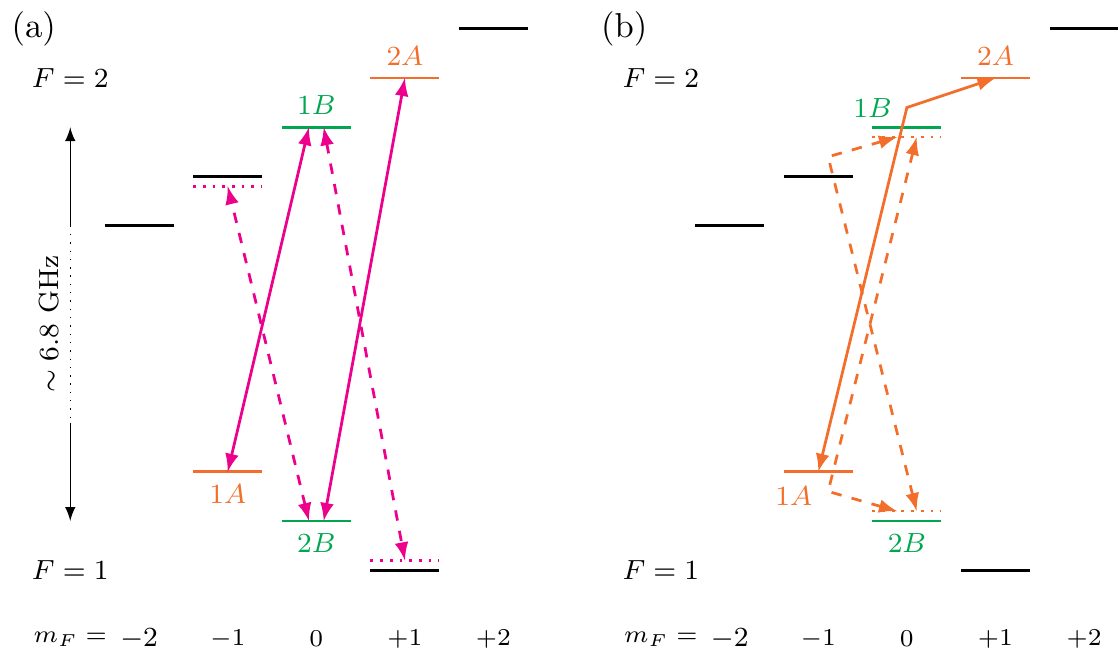}
    \caption{
        Level structure of the \rb{} hyperfine split ground state in the linear Zeeman regime (not to scale).
        The color coding is consistent with Fig.~\ref{experiment}:
        Orange indicates system $ A $, green system $ B $, and magenta the splitting transitions.
        The desired transitions are indicated by solid lines, the undesired close-to-resonance transitions are represented by dashed lines.
        (a) Transitions involved in the splitting of the condensate into systems $A$ and $B$.
        (b) Transitions involved in driving spin rotations of $ \hat{\mathbf{S}}^A $ after splitting. 
    }
    \label{trans_split}
\end{figure*}

In the experimental sequence up until the splitting, the populated states and applied driving frequencies are such that they do not lead to any undesired couplings.
However, the driving fields used for the splitting pulse and for the subsequent rotations of $ \hat{\mathbf{S}}^A $ are close to resonance with other relevant transitions, see Fig.~\ref{trans_split}.
It is possible in principle to suppress the undesired transitions by polarization selection rules, using only driving fields with the correct circular polarization. However, since the condensate is close to the metallic atom chip surface, this is challenging to attain in our experiment and we observe that the driving fields contain all polarization components. We therefore have to ensure the selectivity by careful choice of driving field frequencies and strength.

In the case of the splitting pulse (Fig.~\ref{trans_split}(a)), the desired transition $ {\ket{1A} \leftrightarrow \ket{1B}} $ is close to resonance with $ {\ket{2B} \leftrightarrow \ket{F=2,m_F=-1}} $ ($ 9~\mathrm{kHz} $ detuning) and the desired  $ {\ket{2A} \leftrightarrow \ket{2B}} $ with $ {\ket{1B} \leftrightarrow \ket{F=1,m_F=1}} $ ($ 9~\mathrm{kHz} $ detuning).
These additional close-to-resonant transitions can cause atoms to be transferred from system $ B $ further to states where they are lost during the application of the magnetic gradient pulse due to the different magnetic moments. 
This loss mechanism is mitigated by the fact that states $ \ket{1B} $ and $ \ket{2B} $ are initially unpopulated and become populated only during the splitting $ \pi / 2 $ pulse.
In addition to this, since we only need to transfer half of the population with the splitting pulse, we can choose a detuning to make the undesired transitions further off resonant.
We carefully choose a combination of detuning and small enough Rabi frequency for this pulse for which we observe no losses to the undesired states, corresponding to a detuning of $ 2.2~\mathrm{kHz} $ and a pulse duration $ {t_{\pi / 2}^\mathrm{split} \approx 70 ~ \mathrm{\mu s}} $.

To rotate $ \hat{\mathbf{S}}^A $ after the splitting, the states $ \ket{1A} $ and $ \ket{2A} $ are coupled by a two-photon transition via the intermediate state $ \ket{F=2,m_F=0} $ (see Fig.~\ref{trans_split}(b)). The same driving fields are near resonance with the transition $ {\ket{1B} \leftrightarrow \ket{2B}} $, driven as a two-photon transition with intermediate states $ \ket{F=2,m_F=-1} $ and $ \ket{F=1,m_F=-1} $ ($ 10~\mathrm{kHz} $ detuning).
Since all transitions involved are two-photon transitions with correspondingly weak effective two-photon Rabi frequencies, this two-photon detuning can be exploited to ensure good selectivity of the pulse. The desired transition $ \ket{1A} \leftrightarrow \ket{2A} $ is driven on two-photon resonance with $ \pi / 2 $ time of $ {t_{\pi / 2}^A \approx 960 ~ \mathrm{\mu s}} $, ensuring that the two-photon detuning of the undesired transitions is sufficient to render spurious rotations of $ \hat{\mathbf{S}}^B $ negligible.

\section{Imaging}\label{app_imaging}
We detect the atoms by resonant absorption imaging \cite{ketterle_1999}, which is based on comparing the shadow cast by the atoms in a resonant laser beam (absorption image) with a reference image taken without atoms.
Since the criteria Eq.~\eqref{E EPR} and Eq.~\eqref{E Ent} depend on the atom number, the accurate calibration of the imaging system is crucial.
This is ensured with three types of calibration measurements:
First of all, the conversion from absorbed light to atom number is made independent of the laser intensity following the method described in Ref.\cite{reinaudi_2007}.
Second, the detectivity of the four states is calibrated by ensuring that the detected total atom number is independent of the relative population in the four states when driving Rabi oscillations between them. Last, the total atom number is calibrated by observing the projection noise of a coherent spin state \cite{muessel_2010} prepared in an equal superposition of $ \ket{1A} $ and $ \ket{2A} $. 

Since the calibration relies on projection noise measurements on equal superpositions of the involved states, it is most accurate for the measurements of $\hat{S}_y$ and $\hat{S}_z$ that enter the inferred variances in the EPR and entanglement criteria. In the measurements of $\hat{S}_x$, nearly all atoms occupy one state, and we observe a slight decrease on the order of $ 3 \% $ in the detected total atom number. Since we can exclude atom loss, we attribute this effect to an effective decrease in detectivity due to the increased density of the cloud. To correct for this effect, we 
determine $ \langle\hat{S}_x^{A,B}\rangle  =\langle n_x^{A,B}\rangle \langle N_{y,z}^{A,B}\rangle/2$, i.e.\ we measure the relative atom number ${n_x^{A,B} = (N_{1,x}^{A,B} - N_{2,x}^{A,B}) / (N_{x}^{A,B})}$ corresponding to the interferometric contrast, and multiply by  $\langle N_{y,z}^{(A,B)}\rangle$, the total atom number detected in the corresponding measurements along $y$ and $z$.

In order to minimize detection noise, we choose the region of interest for counting atoms to be as small as possible, while still including nearly all of the atomic signal.
We do this by selecting elliptical regions that include $ \approx 97\% $ of the total detected atom number.
To avoid any artificial noise reduction due to the small discarded signal (see the supplementary materials of Ref.~\cite{fadel_2018} for a discussion of this effect), we perform the atom number calibration after the region of interest has been determined.

Since detection noise is mainly composed of photon shot noise, we can reduce it by choosing longer imaging pulses, as long as atomic diffusion during the imaging pulse doesn't increase the cloud size on the image too much.
We find the lowest imaging noise for an imaging pulse duration of $70 ~ \mathrm{\mu s}$.
During this time, the radiation pressure accelerates the atoms, inducing a relevant Doppler shift.
To maintain the resonance throughout the process, we chirp the frequency of the imaging pulses.

We use a fringe removal algorithm for optimized reference images based on linear combination of actual reference images \cite{ockeloen_2010}. Besides minimizing the effect of interference fringes on the reference images, this procedure further reduces the contribution of photon shot noise of the reference images.

With these optimizations the resulting photon shot noise corresponds to about $\pm 3$~atoms in every state. 

\section{Data analysis}\label{app_data_analysis}
To determine the EPR and entanglement criteria, we take 4500 measurements overall within 42 hours of measurement time. We subdivide the data into blocks for which we evaluate the criteria separately.
Each block consists of hundred $ \hat{S}_z $, hundred $ \hat{S}_y $, and twenty $ \hat{S}_x $ measurements. The measurements along $x$ are performed ten times along the positive and negative spin directions, respectively, to reduce bias from a potential difference in the detectivity. This procedure renders the analysis robust against the effect of small, slow drifts of the experimental conditions during the many hour long measurement run, while the quantum noise of the atoms, which occurs in each shot of the experiment, is unchanged. 
The values of the EPR and entanglement criteria listed in the main text are the averages of the values obtained for the individual blocks.
We verified that analyzing the whole data set in one block does not change the conclusions of our paper, in this case we obtain $\mathcal{E}_\mathrm{EPR}^{A \rightarrow B} = 0.87 \pm 0.04~$ and $ \mathcal{E}_\mathrm{EPR}^{B \rightarrow A} = 0.82\pm 0.04$.

\section{Radio-frequency and microwave signal generation and local oscillator phase}\label{app_rf_mw_phase}
The microwave and radio-frequency driving fields are generated with several commercial function generators.
To allow phase control and simultaneous generation of the two frequencies needed for the splitting of the two-component condensate, we use an IQ -modulated microwave signal (microwave generator: Rohde \& Schwarz SGS100A).
The modulation signal is generated by a two-channel arbitrary waveform generator (Keysight 33522B).
The radio-frequency signal needed for the two-photon transition of system $A$ is generated with a separate radio-frequency source (Photonics Technologies VFG 150).

The sequence of pulses is  generated by programming lists of frequencies, amplitudes, and phases that are executed upon trigger and by gating the respective function generator signals.
The digital signals used for triggering and gating are derived from a common sampling clock.
All function generators are referenced to a GPS-disciplined 10 MHz crystal oscillator (Stanford Research Systems FS752), providing long term frequency stability. However, absolute phase stability is not ensured since not all devices share a common sampling clock. 

In a standard Ramsey experiment, a constant phase offset of the local oscillator does not matter, since the initial coupling of the two states and the final readout of the phase is done with the same local oscillator and an offset of its phase drops out. This scenario applies in our case to the spin $A$, composed of the states $1A$ and $2A$. The two-photon transition between these two states is driven by the microwave generator, the IQ modulator and the RF source. A constant phase offset in any of these devices will not affect the readout of the phase of spin $A$ and thereby $\hat{S}^A_y$. 

For spin $B$ the situation is more complicated. The phase of system $B$, and thereby $\hat{S}^B_y$, is determined by three processes. Initially, spin $B$ inherits the phase of spin $A$, which as above is determined by the phase of the microwave generator, the IQ modulator and the RF source. The second process is the actual splitting in spin space by the two tone microwave signal, which is generated by the microwave generator and the IQ modulator. The same devices are also generating the coupling for the last process, which is the final rotation of spin $B$. In contrast to spin $A$, the RF source only appears in the first process and therefore a phase offset of this source with respect to the other devices does not drop out in the phase sensitive measurement of $\hat{S}^B_y$. 

Since the RF signal and the IQ modulator signal are not derived from the same sampling clock, fluctuations in the relative triggering delay of the devices by $\delta t$ will lead to phase changes of $\omega_\mathrm{RF}\delta t$ in system $B$, where $\omega_\mathrm{RF}\approx 2\pi\times 1.79$ MHz.
We measure typical timing fluctuations of $\delta t$ on the order of 4~ns. The fluctuations are actually small when compared to the large variance observed in the measurement of $ \hat{S}_y^B $, since this is the anti-squeezed direction. However, they are relevant for the value of the much smaller inferred variance. 

These fluctuations are a technical limitation that can be resolved in the future with a different setup of generators for the driving fields.
In the present experiment, we correct them by directly measuring $\delta t$. We determine $\delta t$ in each shot of the experiment from a measurement of the phase of the radio-frequency generator with respect to the starting time of the IQ modulator sequence with an oscilloscope. 
This classical information can then be used for a better estimation of $\hat{S}^B_y= \hat{S}^B_{y,\mathrm{measured}}+g_{\delta t}\delta t$ with an optimized gain parameter $g_{\delta t}$. In Fig.~\ref{spin correlations}, the data points of $\hat{S}^B_y$ are corrected in this way. Furthermore, we use this correction in the evaluation of the EPR criterion Eq.~\eqref{E EPR}. We stress that while this correction based on additional classical information can reduce classical noise, it cannot lead to a violation of the EPR inequality, which can only be achieved by sufficiently strong entanglement between the two systems \cite{reid_2009}.
The correction based on this classical information is not applied in the evaluation of the entanglement criterion.

\end{document}